\documentclass[aps,prb,twocolumn]{revtex4}
\usepackage{graphics,amsmath}
\begin{document}
\title{Aharonov-Bohm oscillations of a particle coupled to
dissipative environments}

\author{F. Guinea}
%\author{\bf DRAFT}
\affiliation{Instituto de Ciencia de Materiales de Madrid,
CSIC, Cantoblanco, E-28049 Madrid, Spain.}

\begin{abstract}
The amplitude of the Bohm-Aharonov oscillations of a particle moving
around a ring threaded by a magnetic flux and coupled to different 
dissipative environments is studied. The decay of the oscillations when
increasing 
the radius of the ring is shown to depend on the spatial features
of the coupling.  When the environment is modelled by the Caldeira-Leggett
bath of oscillators, 
interference effects are suppressed beyond
a finite length, even at zero temperature.
A non trivial renormalization of the Aharonov-Bohm oscillations is
also found when the particle is
coupled by the Coulomb potential to a dirty electron gas.
A finite renormalization
of the Aharonov-Bohm oscillations is obtained for other models
of the environment.
\end{abstract}

\date{\today}
\maketitle

\section{Introduction}
Phase coherence in metallic systems has been extensively studied since
experiments suggested that the dephasing time, $\tau_\phi$, 
seemed to saturate to a constant value at low temperatures\cite{MJW97},
in apparent contradiction with the accepted theory\cite{AAK82,SAI90}.
It has been argued that voltage fluctuations lead to a dephasing time
consistent with saturation at low
temperatures\cite{GZ98,GZS01}, although
related calculations lead to different results\cite{AAV01,KB01}
(see also\cite{B01}).

Comparison between the different calculations of the dephasing time of 
low energy electrons in metals is obscured by the various approximations
required to deal with the interactions and quenched disorder. 
The cause of dephasing, however, is the existence of a dynamic environment
interacting with the electrons. A simpler situation is presented when the
environment is different from the particles whose dephasing is being
studied. Even if that is the case, in a many particle system the
environment induces interactions between the particles. Thus, the
simplest case when dephasing at low temperatures can be studied is that of
a single particle coupled to an external dissipative environment.
The problem can also
be relevant to studies of quantum effects in heavy particles
at metallic surfaces.

We will study here
he amplitude of the Aharonov-Bohm oscillations of
the particle moving around a ring of radius $R$ threaded by a magnetic
flux $\Phi$. This quantity provides information about the
suppression of quantum interference due to the environment.
We will not attempt to define
a dephasing time. On the other hand, 
the dependence of the Aharonov-Bohm oscillations
on the radius of the ring allows us to define, in certain cases,
a length scale, $R_\phi$, beyond which the oscillations
decay exponentially or have a gaussian dependence on the radius. 
 
The simplest quantity which can be studied which depends
on the flux is the free energy. At zero temperature, and in the
absence of dissipative effects, the amplitude of the oscillations
of the energy as function of $\Phi$ is of order $\hbar^2 / ( M R^2)$,
where $M$ is the mass of the particle. The power law dependence of this 
scale on the length of the path of the particle can be interpreted as
the absence of a typical length for the suppression of
quantum coherence effects, at zero temperature.

In the present work, we estimate how this amplitude is changed when
the particle is coupled to a dissipative bath. The next section presents
the model used to analyze this problem. Section III discusses
specific results for five types of dissipative environments: 
i) The Caldeira-Leggett harmonic bath\cite{CL81}
(a closely related aspect of this model was analyzed in\cite{GZ98b}), 
ii)  Dissipation with
a periodic spatial dependence (the dissipative quantum rotor\cite{GS85}),
iii) Dissipation with a gaussian spatial dependence (the DVD 
model\cite{C97}),
iv) Dissipation induced by a local coupling to the low energy modes
of a clean metallic system\cite{G84},  and v) Dissipation induced
by the excitations of a dirty metallic system. Section IV presents the
conclusions.
\begin{figure}
\resizebox{8cm}{!}{\rotatebox{0}{\includegraphics{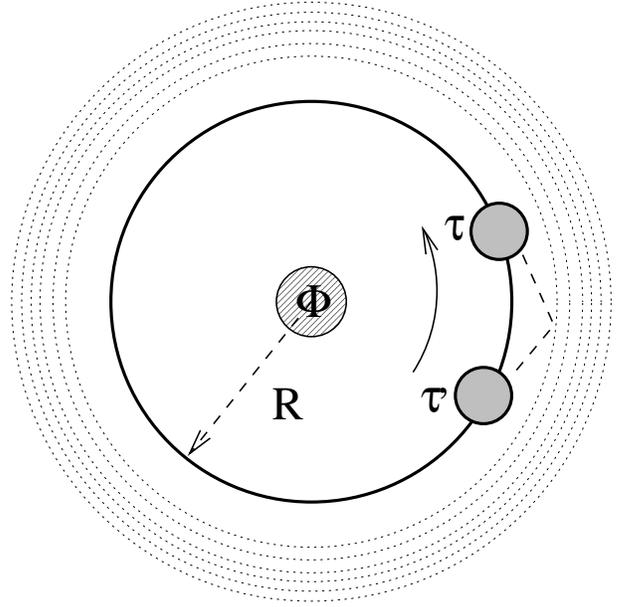}}}
\caption{Schematic picture of the system studied in the text.
A particle interacting with a dissipative environment moves around a
circle of radius $R$. The circle is threaded by a magnetic
flux, $\Phi$.}
\label{fig1}
\end{figure}

\section{The model.} 
We assume that the degrees of freedom in the environment can be
integrated out, leading to retardation effects in the equations
of motion of the particle that we are interested in. These effects
are best studied using the path integral formulation of quantum
mechanics\cite{CL83}. Then, the action associated to each path 
of the particle, in the absence of a magnetic flux,
can be written as:
\begin{equation}
\frac{S}{\hbar} = \int d \tau 
\frac{M}{2 \hbar} \left( \frac{\partial X}{\partial \tau}  \right)^2
+ \int d \tau d \tau' K [ X ( \tau ) - X ( \tau' ) , \tau - \tau' ]
\label{action}
\end{equation}
where $M$ is the mass of the particle, and $X ( \tau )$ is its
position at time $\tau$. The kernel $K ( X , \tau )$ includes the 
information about the environment and its interaction with
the particle. At long times, one has:
\begin{equation}
K ( X , \tau ) \approx \frac{{\cal K} ( X )}{| \tau |^{2}
}
\label{asympt}
\end{equation}
where ${\cal K}$ depends only on the spatial coordinates.

A simple choice is given by the Caldeira-Leggett
model\cite{CL81,CL83}:
\begin{equation}
K ( X , \tau ) = \frac{\gamma}{2 \pi \hbar} \frac{X^2}{\tau^2}
\label{CL}
\end{equation}
and ${\cal K} ( X ) = ( \gamma X^2 ) / ( 2 \pi \hbar )$,
where $\gamma$ is the friction coefficient which describes the dynamics
of the particle when $\hbar \rightarrow 0$. 

Two extensions of the kernel in eq.\{\ref{CL}\} are\cite{GS85,C97}:
\begin{eqnarray}
K ( X , \tau ) &= &\frac{\alpha}{2 \pi} \frac{\sin^2 [ X / (4 \pi L ) ]}{
\tau^2} \nonumber \\
K ( X , \tau ) &= &\frac{\gamma l^2}{2 \pi \hbar} \frac{e^{- X^2 / ( 2 l )}}
{\tau^2}
\label{rotor}
\end{eqnarray}
In the first case, the macroscopic friction coefficient is
$\gamma = ( \hbar \alpha ) / ( 4 \pi L )^2$. In the second case, $l$ is
a length which defines the spatial range of the interactions
mediated by the environment.

When the coupling between the particle and the environment is weak,
one can use perturbation theory. In this work, we will study the 
coupling of the particle to a metallic system by means of a 
local potential of strength $U$ and range $a$:
\begin{equation}
{\cal H}_{int} = U \int d^3 {\bf \vec{r}} {\cal F} \left( 
\frac{| {\bf \vec{r}} - {\bf \vec{R}} |}{a} \right) \rho ( {\bf \vec{r}} )
\end{equation}
where ${\bf \vec{R}}$ is the coordinate of the particle,
$\rho ( {\bf \vec{r}} )$ is the density operator
of the electron gas, and ${\cal F} ( u ) \sim 0$ if $u \gg 1$.
This coupling leads to a retarded interaction along the path
taken by particle. Using second order perturbation theory,
this interaction can be cast as\cite{G84}:
\begin{widetext} 
\begin{equation}
S_{int} = \frac{U^2}{2 \pi \hbar} 
\int d \tau 
d \tau'  
d^3 {\bf \vec{r}}
d^3 {\bf \vec{r}}'  d \omega  d^3 {\bf \vec{k}}
{\cal F} \left(
\frac{| {\bf \vec{R}} ( \tau ) - {\bf \vec{r}} |}{a} \right)
{\cal F} \left(
\frac{| {\bf \vec{R}} ( \tau' ) - {\bf \vec{r}}' |}{a} \right)
e^{i {\bf \vec{k}} ( {\bf \vec{r}} - {\bf \vec{r}}' )}
e^{i \omega ( \tau - \tau' )} \chi ( {\bf \vec{k}} , \omega )
\label{Sint}
\end{equation}
\end{widetext}
where $\chi ( {\bf \vec{k}} , \omega )$ is the density-density
correlation function of
the metal. 

The coordinate $X$ of the particle will be restricted to lie
within a circle of radius $R$, so that the motion can be  also
described in terms of the angle $\theta ( \tau )$, as schematically
shown in Fig.[\ref{fig1}]. In terms of
this angle, we have:
\begin{equation}
\left| {\bf \vec{R}} ( \tau ) - {\bf \vec{R}} ( \tau' ) \right| = 2 R \sin
\left[ \frac{ \theta ( \tau ) - \theta ( \tau' )}{2} \right]
\end{equation}
The action, eq.\{\ref{action}\}, can be expanded in circular harmonics as:
\begin{widetext}
\begin{equation}
\frac{S}{\hbar}
= \int d \tau \frac{M R^2}{2 \hbar} \left( \frac{\partial \theta}
{\partial \tau} \right)^2 + \int d \tau d \tau' \sum_n
\alpha_n \frac{\sin^2 \left\{ \frac{n [ \theta ( \tau  ) -
\theta ( \tau' ) ]}{2} \right\} }{| \tau - \tau' |^{2}}
\label{periodic}
\end{equation}
\end{widetext} 
where the $\alpha_n$'s are dimensionless constants,
given by:
\begin{equation}
\alpha_n = \frac{1}{2 \pi} \int d \theta e^{i n \theta} {\cal K}
\left[ 2 R \sin ( \theta / 2 ) \right]
\label{Fourier}
\end{equation}
The action in eq.\{\ref{periodic}\}
resembles closely the quantum rotor model\cite{K77} studied extensively in
relation to Coulomb blockade in normal tunnel junctions\cite{GS85}. 
In the present case, the short range interaction is related to the
energy scale $\hbar^2 / ( 2 M R^2 )$. When studying the quantum rotor model in 
the context of mesoscopic junctions, this scale corresponds, 
at the same time, to the charging energy of the junction, $E_C$, and to the
upper cutoff in the spectrum of the environment coupled to the variable
under study. In the present case, the equivalent to $E_C$ depends on the 
radius of the ring, $R$, and the two scales should be kept separate.
Thus, and following the conventional notation for tunnel junctions,
the model in eq.\{\ref{periodic}\} contains two scales, $E_C =
\hbar^2 / ( 2 M R^2 )$, and an
energy, $\Lambda_0$, which defines
the short time cutoff in the kernel in eq.\{\ref{asympt}\}.
We will assume that $E_C \ll \Lambda_0$.

In order analyze the problem, we extend the Renormalization Group 
approach initially discussed in\cite{K77} to the action in
eq.\{\ref{periodic}\}. We need to consider the scaling of
the parameters $\{ \alpha_n \}$, and, for completemess,
we also consider the renormalization of the dimensionless
coupling $\tilde{E}_C = E_C / \Lambda$ due to the high energy
excitations of the environment. 
We lower the effective high energy cutoff
from $\Lambda$ to $\Lambda - d \Lambda$, and rescale the dimensionless
parameters $\tilde{E}_C$ and 
$\{ \alpha_n \}$. Ths leads to the equations:
\begin{eqnarray}
\frac{\partial \tilde{E}_C}{\partial l} &= &\tilde{E}_C - c_1
\sum_n n^2 \alpha_n \tilde{E}_C^2 \nonumber \\
\frac{\partial \alpha_n}{\partial l} &= &-\frac{1}{2 \pi^2}
\frac{n^2 \alpha_n}{c_2 \tilde{E}_C^{-1} + \sum_m m^2 \alpha_m}  
\label{RG}
\end{eqnarray}
where  $l = - \log ( \Lambda )$ and $c_1$
are $c_2$ are constants of order unity.
The scaling equation for $\tilde{E}_C$ ceases to be valid when
$\tilde{E}_C \sim 1$. The renormalization of the 
$\{\alpha_n\}$'s is only significant when
$\tilde{E}_C \sim 1$. In addition, the equations for the $\{ \alpha_n \}$'s
must be halted when
$\sum_n n^2 \alpha_n \sim 1$.
We can write the second equation in \{\ref{RG}\} as:
\begin{equation}
\frac{\partial \sum_n n^2 \alpha_n}{\partial l} = - \frac{1}{2 \pi^2}
\frac{
\sum n^4 \alpha_n}{c_2 \tilde{E}_C^{-1} + \sum n^2 \alpha_n} 
\label{sum}
\end{equation}

The equation which determines the flow of $\tilde{E}_C$ shows two
regimes, depending on whether  
\begin{equation}
\kappa = \frac{( \sum_n n^2 \alpha_n^0 )^2}{ \sum n^4 \alpha_n^0} 
\tilde{E}_C^0
\label{kappa}
\end{equation}
is smaller or greater than one. When $\kappa \ll 1$, the absolute value
of $E_C$ is not changed by the renormalization of the modes
with energies between $\Lambda_0$ and $E_C$ itself. In this case,
$E_C$ defines the cutoff of an effective theory where the only couplings
left are the $\{\alpha_n\}$'s, in a similar fashion to the usual
case studied in\cite{K77}. This is the situation which is most relevant
to the calculations to be performed in the following. 
The effective low energy scale is the given by:
\begin{equation}
E_C^{ren} \sim E_C \exp \left[ -2 \pi^2 \frac{(  \sum n^2 \alpha_n )^2}
{\sum n^4 \alpha_n} \right]
\label{ren}
\end{equation}

If $\kappa \gg 1$,
the renormalization of $\tilde{E}_C$ is determined by the coupling
to the environment, and only when this coupling flows towards
zero $\tilde{E}_C$ can approach unity. 
the scale at which $\tilde{E}_C \sim 1$ implies that:
\begin{equation}
E_C^{ren} \sim \Lambda_0 e^{- \tilde{E}_C^0 \kappa} e^{- \kappa } 
\sim \Lambda_0 e^{- \lambda / \tilde{E}_C^0}
\label{eren2}
\end{equation}
where $\lambda$ is a contant.
   
The contribution to the 
action from the magnetic flux threading the circle is a topological
term, which can be written as:
\begin{equation}
\frac{\delta S}{\hbar} =  
i  \frac{\Phi}{\Phi_0} \int d \tau \frac{\partial \theta}
{\partial \tau}
\label{topological}
\end{equation}
where $\Phi_0$ is the quantum unit of flux.
This term enters in the effective action in the same way a gate voltage
is included in the correponding problem of tunneling in a small
junction, which has also been studied in the literature\cite{HSZ99,BHGS00}.
It is known that the fluctuations in the free energy at low temperatures
are determined by the renormalized value of $E_C^{ren}$. In the following,
we will assume that the amplitude of the Aharonov-Bohm oscillations in
the free energy of a particle moving around a ring are also determined
by $E_C^{ren}$, as calculated using the scaling equations \{\ref{RG}\}.
This scale is given in eq.\{\ref{eren}\} or eq.\{\ref{eren2}\}, where
$E_C^0 \approx \frac{\hbar^2}{M R^2}$. 

\section{Results.}
\subsection{Caldeira-Leggett model.}
The dynamics of a quantum particle around a ring, using the Caldeira-Leggett
bath of oscillators as a model for the environment, has been
considered in\cite{GZ98b}. The analysis presented here of
the Aharonov-Bohm oscillations is consistent with the results in\cite{GZ98b}.
 
For the present model,
the coefficients in the harmonic expansion in
eq.\{\ref{periodic}\} reduce to:
\begin{equation}
\alpha_n = \delta_{n,1} \frac{\gamma R^2}{\hbar} = \alpha
\end{equation}
and the parameter in eq.\{\ref{kappa}\} is
$\kappa \sim ( \hbar \gamma ) / ( M \Lambda_0 ) \ll 1$.
We recover the quantum rotor in its simplest version.
Using eq.\{\ref{ren}\}, we find:
\begin{equation}
E_C^{ren} \sim E_C e^{- 2 \pi^2 \alpha} \sim 
\frac{\hbar^2}{M R^2} e^{- ( 2 \pi^2 \gamma R^2 )
/ \hbar }
\end{equation}
The Aharonov-Bohm  oscillations 
will show a gaussian decay as the radius of the ring
is increased. Thus, quantum interference effects are suppressed
beyond certain length,
$R_\phi \sim \sqrt{( \hbar / \gamma )}$.
This suppression of the Aharonov-Bohm oscillations are
in qualitative agreement with the vanishing of the Landau
diamagnetism of a particle interacting with a Caldeira-Leggett
bath of oscillators in a magnetic field\cite{DS97}, at zero
temperature. In the language used in\cite{DS97}, our results
suggest that, in a conatant magnetic field $B$, the properties of
the system at zero temperature are a function of the dimensionless
ratio $( \gamma r_c^2 ) / \hbar = ( \gamma c ) / ( e B )$,
where $r_c$ is the cyclotron radius. This gaussian suppression of 
the Aharonov-Bohm oscillations at zero temperature is consistent
with the similar suppression of interference effects between time reversed
paths discussed in\cite{GZ98b}.

Note that, as $\alpha \sim O ( R^2 )$, the scaling equations, \{\ref{RG}\}
are valid for large values of $R$.
\subsection{The dissipative quantum rotor.}
We now consider the retarded interaction described in eq.\{\ref{rotor}\}.
The spatial dependence of the kernel allows for only one Fourier component,
so that the decomposition needed in eq.\{\ref{periodic}\} becomes:
\begin{equation}
\alpha_n =  \alpha \delta_{n,n_0} 
\end{equation}
where $\alpha$ is the dimensionless constant
in eq.\{\ref{rotor}\}, and $n_0 = R / L$, where $L$ 
is the period of the kernel.
The parameter in eq.\{\ref{kappa}\} is
$\kappa \sim ( \alpha   \hbar^2 ) / ( M L^2 \Lambda_0 )  \ll 1$.

Using eq.\{\ref{ren}\}, we find:
\begin{equation}
E_C^{ren} \sim 
\frac{\hbar^2}{M R^2}
 e^{- 2 \pi^2 \alpha } 
\end{equation}
Hence, the Aharonov-Bohm oscillations show the
dependence on $R$ as in
the non dissipative case, although the coupling to the environment
leads to a finite reduction of the amplitude.
\subsection{Coupling via a kernel with a gaussian dependence on distance.}
This model\cite{C97} is defined in terms of a kernel, eq.\{\ref{asympt}\},
\begin{equation}
{\cal K} ( X ) = \frac{\gamma l^2}{2 \pi} e^{- X^2 / ( 2 l^2 )}
\end{equation}
where $\gamma$ is the friction coefficient, and $l$ is a length which 
characterizes the spatial correlations in the bath.

We now use $X = 2 R \sin ( \theta / 2 )$, and perform the Fourier
transform in eq.\{\ref{Fourier}\} using the saddle
point approximation, to obtain:
\begin{equation}
\alpha_n \approx \frac{\gamma l^2}{\hbar}
\left( \frac{l}{R} \right) e^{- ( n^2 l^2 ) / ( 2 R^2 )}
\end{equation}
The parameter $\kappa$ in eq.\{\ref{kappa}\} is:
\begin{equation}
\kappa \sim \frac{\gamma l^2}{\hbar} \frac{\hbar^2 \Lambda_0}{M R^2}
\end{equation}
where $\Lambda_0$ is the high energy cutoff of the environment.
We have $\kappa \ll 1$ when $R \gg l$. The renormalization of
the Aharonov-Bohm oscillations is given by:
\begin{equation}
E_C^{ren} \sim \frac{\hbar^2}{M R^2} e^{- ( 2 \pi^2 \gamma l^2 ) / \hbar}
\end{equation}
The amplitude of the Aharonov-Bohm oscillations is reduced by a finite
factor.
\subsection{Coupling to a clean electron gas.}
The density-density correlation function
of a three dimensional electron gas can approximately
be written, at low frequencies, as:
\begin{equation}
\chi ( {\bf \vec{k}} , \omega ) \approx \frac{|\hbar \omega |}
{2 \pi^2 (\hbar^2 / m)^2
| {\bf \vec{k}} | } \theta \left( 1 -
\frac{| {\bf \vec{k}} |}{2 k_F} \right)
\label{susc}
\end{equation} 
where $m$ is the electron mass, $k_F$ is the Fermi 
wavevector, and $\theta ( u ) $ is the
step function. The high energy cutoff
is $\Lambda_0 \sim E_F \sim ( \hbar^2 k_F^2 ) / (2 m )$, where $E_F$
is the Fermi energy.
The interaction term in the effective action, eq.\{\ref{Sint}\},
can be written as (see eq. \{\ref{asympt}\}):
\begin{equation}
S_{int} \approx \frac{1}{2 \pi \hbar} \int
d^3 {\bf \vec{R}} ( \tau ) d^3 {\bf \vec{R}} ( \tau' )
\frac{{\cal K} [ | {\bf \vec{R}} ( \tau )  - {\bf \vec{R}} ( \tau' ) | ]}
{( \tau - \tau' )^2}
\end{equation}
where:
\begin{equation}
{\cal K} ( | {\bf \vec{R}} | ) \approx  \frac{ ( U a^3 )^2 }
{2 \pi^2 ( \hbar^2 / m )^2}
\frac{ 1 - \cos ( 2 k_F | {\bf \vec{R}} | ) }{| {\bf \vec{R}} |^2}
\end{equation}
and we are assuming that $k_F a \gg 1$. 
%Inserting the interaction potential, the coefficients in the
%expansion in eq.\{\ref{periodic}\} can be written as:
%\begin{widetext}
%\begin{equation}
%\alpha_n =  
%\frac{2}{\pi} \frac{ ( U a^3 )^2}{( \hbar^2 / m )^2 } 
%\int d \theta e^{i n \theta}
%\frac{1 - \cos [ 2 k_F | 2 R \sin ( \theta / 2 ) | ]}
%{  [ 2 R \sin ( \theta / 2 ) ]^2}
%\end{equation}
%\end{widetext}
We perform the $\theta$ integration
in eq.\{\ref{Fourier}\} using the saddle point approximation,
and we obtain:
\begin{equation}
\alpha_n \sim 
\frac{( U a^3 )^2 k_F^2}{( \hbar^2 / m )^2 ( k_F R )}
e^{- ( 3 n^2 ) / ( k_F R )^2}
\label{integral}
\end{equation}
and, finally, we can approximate:
\begin{equation}
\alpha_n \sim \frac{( U a^3 )^2 k_F}{( \hbar^2 / m )^2 R} 
\sim \frac{\delta}{k_F R} \, , \, \, \, \, \, \, \, \, n \ll k_F R
\label{alpha}
\end{equation}
where we are defining the phaseshift $\delta$ induced by the particle
on the states near the Fermi level of the electron gas as:
\begin{equation}
\delta \sim \frac{ ( U a^3 )^2 k_F^6}{E_F^2}
\label{phaseshift}
\end{equation}

Using eq.\{\ref{alpha}\}, we obtain:
\begin{eqnarray}
\sum_n n^2 \alpha_n &\sim & \delta ( k_F R )^3 \nonumber \\
\sum_n n^4 \alpha_n &\sim &\delta ( k_F R )^5 
\end{eqnarray}
and the parameter in eq.\{\ref{kappa}\} is $
\kappa \sim ( \hbar^2 k_F^2 / ( M E_F ) \ll 1$.
 
Using equation \{\ref{ren}\}, we find:
%\begin{widetext}
\begin{equation}
E_C^{ren} \sim \frac{\hbar^2}{M R^2}
e^{- 2 \pi^2 \delta}
\label{eren}
\end{equation}
%\end{widetext}
Hence, the amplitude of the Aharonov-Bohm, as in the quantum rotor
case, is reduced by a finite factor as $R \rightarrow \infty$. 
Similar results can be obtained using the response function of
the two dimensional electron gas. The qualitative features of
the RPA response function used here are generic to the response
of a clean electron gas, whose low energy excitations can be
described in terms of Landau's theory.
Finally, the results presented in this subsection
remain valid if the local coupling between the particle and 
the electron gas is be replaced by a screened electrostatic potential.
If we assume that the charge of the particle is $e^*$ and the
charge of the electrons is $e$,  the expression $U a^3$ in
eq.\{\ref{phaseshift}\} has to be replaced by $e^* e k_{FT}$, 
where $
k_{FT} = \sqrt{( 4 e^2 m k_F ) / ( \pi \hbar^2 )}$ is the Fermi-Thomas 
wavevector.

\subsection{Coupling to a dirty electron gas.}
In this case, the susceptibility is given by:
\begin{equation}
\chi ( {\bf \vec{k}} , \omega ) \approx 
\nu  \frac{D | {\bf \vec{k}} |^2}{i \omega + D | {\bf \vec{k}} |^2}
\label{susc2}
\end{equation}
where $D$ is the diffusion coefficient, $D \sim v_F l \sim
( \hbar k_F l ) / m$, $l$ is the mean free path, and $\nu$
is the density of states. 
This expression is valid for $| {\bf \vec{k}} | \ll l^{-1}$,
and $\omega \ll \Lambda_0 \sim ( \hbar D ) / l^2$.
In the following, we consider separately the case when the
coupling between the particle and the electrons is by means of
a local potential, as generically described in eq.\{\ref{Sint}\},
or by a screened Coulomb potential. Unlike in the case of
a clean electron gas, discussed in the previous subsection,
the two situations are not equivalent.  

i) Coupling by a short range potential.

The time dependence of this kernel is not a simple power law.
The function $K ( X , \tau )$ in eq.\{\ref{Sint}\} becomes:
\begin{equation}
K ( X , \tau ) \sim \nu ( U a^3 )^2\frac{
-D \tau + X^2}{\hbar \sqrt{D \tau}  D^2 \tau^3}
e^{- X^2 / ( D \tau )}
\label{kdirty}
\end{equation}
We can now take $X = 2 R \sin ( \theta / 2 )$, and decompose
\{\ref{kdirty}\} in circular harmonics. Using the saddle point
approximation to perform the integral over $\theta$, we
obtain:
\begin{widetext}
\begin{equation}
K_n ( \tau ) \sim \frac{\nu ( U a^3 )^2}{\hbar D R}
\left( \frac{1}{\tau^2} + \frac{n^2 D}{R^2 \tau}
\right) e^{-
( D \tau n^2 ) / R^2} \sim
\frac{\delta}{k_F^2 R l}
\left( \frac{1}{\tau^2} + \frac{n^2 D}{R^2 \tau}
\right) e^{-
( D \tau n^2 ) / R^2} 
\label{K_n}
\end{equation}
\end{widetext}
We have defined, as in the previous subsection, the phaseshift
as $\delta \sim ( U / E_F )^2 ( k_F a )^6$ and
$\nu \sim k_F^3 / E_F$.
The terms in eq.\{\ref{K_n}\} decay exponentially for values of $\tau$
larger than $E_T^{-1} \sim [ ( \hbar D ) / R^2 ]^{-1}$, where 
$E_T$ can be defined as the Thouless energy for the
electrons moving around paths comparable to the ring.
Hence, in the more physical regime, 
$E_T \gg E_C = \hbar^2 / ( M R^2 )$, or, alternatively,
$( k_F l ) ( M / m ) \gg 1$, the effect of the environment
is exponentially suppressed, and there will be no 
significant renormalization of the Aharonov-Bohm oscillations.

ii) Coupling by the Coulomb potential.

In this case, one has to replace the product 
$( U a^3 )^2 \chi ( {\bf \vec{k}} , \omega )$ used previously by:
\begin{equation}
{\rm Im} \left\{ \frac{4 \pi e e^*}{| {\bf \vec{k}} |^2 +   4 \pi e e^*
\chi ( {\bf \vec{k}} , \omega )} \right\} \approx
\frac{| \omega |}{| {\bf \vec{k}} |^2 D \nu}
\label{kernel_dirty}
\end{equation}
where $e^*$ is the charge of the particle, and $e$ that of the electrons.
The kernel which describes the retarded interactions
decays as $\tau^{-2}$ at long times.
The spatial dependence of the kernel ${\cal K}$, as defined in
eq.\{\ref{asympt}\}, becomes:
\begin{equation}
{\cal K} ( X ) \approx \int_{| {\bf \vec{k}} |
\ll l^{-1}} d^3 {\bf \vec{k}}
\frac{\sin ( | {\bf \vec{k}} | X )}
{\hbar D \nu | {\bf \vec{k}} |^3 X}
\end{equation}
Setting $X = 2 R | \sin ( \theta ) / 2 |$, and performing the
Fourier tranform defined in eq.\{\ref{Fourier}\}, we obtain:
\begin{widetext}
\begin{equation}
\alpha_n \sim 
\int_{| {\bf \vec{k}} |
\ll l^{-1}} d^3 {\bf \vec{k}}
\frac{1}{| {\bf \vec{k}} |^3 R}
e^{- n^2 / ( | {\bf \vec{k}} | R )^2}
\sim \int_{k \sim n / R}^{k \sim 1 / l} \frac{d k}{k} \sim
\left\{ \begin{array}{lr}
\frac{1}{\hbar \nu D R} \log \left( \frac{R}{n l} \right) &n \ll R / l \\
0 & n \gg R / l
\end{array} \right.
\end{equation}
\end{widetext}
The parameter $\kappa$ defined in eq.\{\ref{kappa}\} is
$\kappa \sim ( M R ) / ( \hbar \nu D l^2 ) \gg 1$. The renormalization
of $E_C$ is, in this case:
\begin{equation}
E_C^{ren} \sim \frac{\hbar^2}{M R^2} 
\left( \frac{l}{R} \right)^{c / ( k_F l )^2}
\end{equation}
where $c$ is a constant of order unity.
%The $R$ dependence of the Aharonov-Bohm oscillations is a power law,
%although with an enhanced exponent than in the case of
%a short range coupling. 
\section{Conclusions.}
We have analyzed the Aharonov-Bohm oscillations in
five specific models for a particle coupled to
different models of dissipative baths.
In most cases, these oscillations are suppressed by a factor which can be
written as $e^{ - c ( \gamma l^2 ) / \hbar}$, where $\gamma$ is the 
macroscopic friction coefficient, $l$ is a length which characterizes
the spatial range of the interactions induced
by the environment, and $c$
is a constant. This factor is independent of the radius of the orbit, $R$.
This is the case, for instance, when the
particle is coupled
to a clean electron gas, where $l \sim k_F^{-1}$.
 
However, when
the environment is the Caldeira-Leggett bath of
oscillators, the renormalization of the
amplitude of the oscillations has a
gaussian dependence on the radius of the circle in which
the particle moves, and the suppression factor mentioned in the
previous paragraph becomes $e^{- c ( \gamma R^2 ) / \hbar}$. 
Hence, quantum interference effects
become negligible beyond a certain length, $R_\phi \sim
\sqrt{\hbar / \gamma}$, where $\gamma$ is the friction coefficient.
A less divergent suppression is also found for a charged particle
coupled to a dirty electron gas, where the dependence of
the Aharonov-Bohm amplitudes is a power law, although different
from the value obtained  in the absence of the environment.
   
This diverse behavior in different models arises from
the spatial range of the retarded interaction induced by the environment.
This difference is lost in the classical limit, which is 
attained at sufficiently high temperatures. 
When the thermal length, $L_T \sim \sqrt{\hbar^2 / ( M T )}$, is much 
shorter than the range of the retarded interaction, the effects of
the environment can be expressed in terms of an
effective friction coefficient, provided that the 
interaction in time decays as $\tau^{-2}$ at zero 
temperature, as in most cases considered here.

The suppression of interference effects, when it exists,
is due to the formation of a screening cloud around the particle,
with contributions from the high energy modes of the environment.
The effect can be cast in terms of the existence of a Franck-Condon
overlap factor which suppresses quantum interference effects\cite{Y01}.
This factor can depend on the length of the path of
the particle around the magnetic flux, leading to the 
suppression of the Aharonov-Bohm oscillations. This interpretation
is consistent with the fact that the same renormalization enters
in the effective mass of the particle.
The conductance can be defined 
as a function of the sensitivity of the ground state
energy to a magnetic flux\cite{K64}. Hence, the Franck-Condon factor
discussed here also reduces the conductance. 
%Note, however, that
%the divergence of the effective mass discussed nere does not imply
%the localization of the particle\cite{SG87}.

Finally, it is worth noting that
the divergence of  the renormalization of the effective mass which 
appears with the suppression of the Aharonov-Bohm oscillations
implies a qualitative change in the propagator of the particle.
The ^^ ^^ quasiparticle peak " at zero momentum, which characterizes
the propagator of a  free particle in the ground state,
is replaced by an incoherent background.
\section{Acknowledgements.}
I am thankful to Y. Imry, R. Jalabert, G. Sch\"on and A. Zaikin
for helpful conversations, and to A. Kamenev for pointing out to
a mistake in an earlier version of the manuscript.
This work was done while at the
Institute for Theoretical Physics, Santa Barbara.
This research was supported in part by the National Science
Foundation under Grant No. PHY99-07949, and MEC (Spain) under
Grant No. PB96/0875.
%\end{multicols}

\end{document}